\documentclass[11pt,amssymb,prl,aps,floatfix,floats,preprintnumbers,showpacs,showkeys,twocolumn,superscriptaddress,groupedaddress]{revtex4-1}
\usepackage{dcolumn}
\usepackage{bm}
\usepackage{amssymb}
\usepackage{color}
\usepackage{graphicx}
\usepackage{graphics}
\usepackage[normalem]{ulem}
\usepackage{chngcntr}
\usepackage{natbib}
\usepackage{amsmath}
\begin{document}
\title{\sc critical acoustics and singular bulk viscosity of quark matter}
\thanks{Presented at the Rencontres de Moriond 2018 on QCD and High Energy Interactions}
\author{Boris O. Kerbikov$^{1}$}
\email[E-mail me at:  ]{bkerbikov@gmail.com}
\affiliation{$^{1}$Alikhanov Institute for Theoretical and Experimental Physics, B. Cheremushkinskya 25, 117218 Moscow, Russia}
\date{\today}
\begin{abstract}
We examine the behavior of sound attenuation and bulk viscosity near the 2'nd order QCD phase transition at finite density. A dynamical model is presented describing the coupled evolution of sound mode and a slow mode of fluctuations.  
\end{abstract}
\pacs{}
\keywords{}
\maketitle
\section{Introduction}
The main result of heavy ion experiments performed over almost two last decades at RHIC and at RHIC and LHC is a discovery of a new form of matter with properties markedly different from the pre-RHIC era predictions. Theoretical efforts accompanying the experimental discoveries added a lot to our understanding of the quark matter properties. There is, however, an unfortunate aspect in the theoretical studies. Only zero, or very small $\mu_B$ domain of the QCD phase diagram in the $(\mu_B,T)$ plane is accessible to lattice Monte-Carlo simulations since for $\mu\neq 0$ the fermion determinant is no longer real.

The analysis of the quark matter properties can be carried out testing its response to external exitations. Transport coefficients characterize how small perturbations from equilibrium are transmitted through the medium. Shortly after the collision and until hadronization the created matter is in the form of an almost ideal liquid. Sound is the only long-lived propagating mode in a near-ideal fluid. Viscosity leads to the dissipation of the sound wave energy. Shear viscosity $\eta$ dominates the sound attenuation in the regime of weak interaction when the system is approximately scale invariant (Stokes absorption). If the medium is compressible and close to deconfiment/confinement transition the bulk viscisity $\zeta$ is responsible for the sound absorption. Close to the critical temperature $T_c$ the sound attenuation is anomalously strong \cite{1,2,3,4,5}. The bulk viscosity $\zeta$ sharply rises towards singalarity near $T_c$ \cite{6,7,8,9,10}. The speed of sound which is by the fluctuation-dissipation theorem related to the sound absorption attains its minimum near $T_c$. This is the softest point of EoS. In ultra-relativistic and ultra-nonrelativistic limits when the trace of the energy-momentum tensor depends only on the energy and particle densities $\zeta = 0$ \cite{11}. As shown by Mandelshtam and Leontovich (ML) long ago \cite{12} and later confirmed by Tisza \cite{13} the anomalous behavior of the sound absorption coefficient $\gamma$ near $T_c$ is caused by the contribution from the slow mode with a large relaxation time $\tau = \Gamma^{-1}$. The divergent bulk viscosity has the same origin. The essence of ML theory is the following \cite{1,2,12,14}. Propagation of the sound wave changes the temperature in compression-rarefaction regions. The slow mode (chemical reactions in the original ML work \cite{12}) can not keep up with this process. During the slow equilibration energy dissipation takes place resulting in anomalous sound absorption and enhanced bulk viscosity. In \cite{14} the sound attenuation near the 2'nd order transition point has been investigated along the lines of the ML theory. In this case the slow mode corresponds to the relaxation of the order parameter which exhibits fluctuations near $T_c$. Let us present a compendium of the results of the ML theory for the low frequency $\omega \tau \lesssim 1$ region \cite{1,2,9,14}

\begin{equation}
\gamma(\omega) = \dfrac{\omega^2 \tau}{2 c_0^3} \dfrac{(c^2_{\infty}-c^2_0)}{1+\omega^2\tau^2},
\label{01}
\end{equation}

\begin{equation}
\zeta(\omega) = \dfrac{\zeta(0)}{1+\omega^2\tau^2}=\dfrac{\varepsilon\tau (c^2_{\infty}-c^2_0)}{1+\omega^2\tau^2},
\label{02}
\end{equation}

\begin{align}
c^2(\omega) = c^2_0 &+ \dfrac{{\omega^2 \tau^2}}{1+\omega^2\tau^2}(c^2_{\infty}-c^2_0) \notag \\ 
&-i \dfrac{{\omega \tau}}{1+\omega^2\tau^2}(c^2_{\infty}-c^2_0).
\label{03}
\end{align}

Here $c^2_0=\left( \dfrac{\partial p}{\partial \varepsilon} \right)_s$ is the ordinary hydrodynamic speed of sound, $c^2_{\infty}=\left( \dfrac{\partial p}{\partial \varepsilon} \right)_{\varphi}$ -- the speed of sound at a given value of the order parameter $\varphi$, $c_{\infty} > c_0$. Expression for $\zeta (0)$ in (\ref{02}) is the known Landau-Khalatnikov formula \cite{14}. In order to equilize the short and long waves distortions caused by the attenuation one introduces the dimensionless attenuation per wavelength $\alpha_{\lambda}=\lambda \gamma$. According to (\ref{01})-(\ref{02}) 

\begin{equation}
\alpha_{\lambda} = \omega\dfrac{\pi \zeta(0)}{\varepsilon c_0^2},
\label{04}
\end{equation}

This relation is used to determine the bulk viscosity of conventional materials from the sound absorption. From (\ref{01}) it follows that $\omega \tau \ll 1$ means $\alpha_{\lambda} \ll \pi$, i.e., weak absorption. The clear-cut picture of the high frequency $\omega \tau \gg 1$ transport coefficients is far from being complete. According to \cite{5} with $\omega \tau$ increasing $\alpha_{\lambda}(\omega)$ slowly approaches the constant value defined by the critical indices. 

The most difficult problem within the slow relaxation ML theory is the behavior of $\zeta$ and $\alpha_{\lambda}$ in the immediate vicinity of the 2'nd order transition temperature, i.e., at $|t| \lesssim Gi$, where $t = (T-T_c)/T_c$, and $Gi$ is the Ginzburg-Levanyuk number. The subtle point is the distinction between hydrodynamic and non-hydrodynamic sectors \cite{2,9,15}. Following \cite{7,8,10} we consider $\omega \to 0$, $T \to T_c$. In various approaches, like $d=4-\varepsilon$ renormalization, modes coupling theory, or isomorphism between the fluid and 3d Ising system the bulk viscosity $\zeta(\omega \to 0,\,T \to T_c)$ shows a power divergence    

\begin{equation}
\zeta \sim \xi^{z-\alpha/\nu} \sim t^{-z\nu+\alpha},
\label{05}
\end{equation}

\begin{figure}[h]\centering
\begin{center}
\resizebox{0.80\columnwidth}{!}{\includegraphics[scale=0.9]{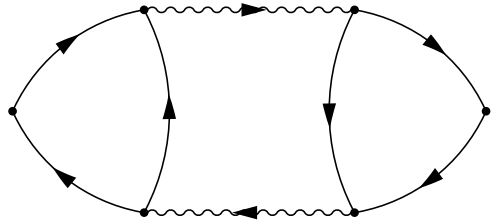}}
\end{center}
\caption{\label{fig:FIG1}Feynman diagram for the AL polarization operator. The way lines correspond to $L(\mathbf{q},\omega)$.}
\end{figure}

Here $\xi$ is the correlation length, $z \simeq 3$ is the dynamical critical exponent, $\nu \simeq 0.6$ is the correlation length critical exponent, $\xi(t)\simeq t^{-\nu}$, $\alpha \simeq 0.11$ is the critical exponent of the heat capacity. According to (\ref{04}) $\alpha_{\lambda}$ diverges near $T_c$ as $\alpha_{\lambda}\sim t^{-1.69}$. The above anomalies of $\zeta$ and $\alpha_{\lambda}$ arise as a result of coupling between the hydrodynamical modes and the slow order parameter mode. Below we present a dynamical model \cite{16} which explicitly contains the slow mode and gives the critical behavior of $\zeta(T)$ very close to (\ref{05}). Consider a region of the QCD phase diagram corresponding to the onset of the 2SC superconductivity. For the orientation purposes one can take $T_c \simeq 40$ MeV, $\mu$(quark) $\simeq 400$ MeV. In color superconductor the exceedingly narrow BCS fluctuation region is replaced by a wide and physically important one $\delta T/T \simeq 10^{-4}$ instead of $(10^{-12}-10^{-14})$ in BCS. When the temperature approaches $T_c$ from above precursor quark pairing takes place. Fluctuations of the pair field are described by a collective mode propagator, or the fluctuation propagator (FP) $L(\mathbf{q},\omega)$ \cite{4}. In \cite{16,17} the FP was derived for the relativistic quark system using either Dyson equation, or the time-dependent Landau-Ginzburg equation with stochastic Langevin forces. The FP reads

\begin{equation}
L(\mathbf{q},\omega)= -\dfrac{1}{\nu}\cdot\dfrac{1}{t+\frac{\pi}{8T_c}(-i\omega + Dq^2)},
\label{06}
\end{equation}

where $\nu=p_0\mu/2\pi^2$, $p_0$ is the Fermi momentum, $D$ is the diffusion coefficient. At small $\omega$ and $q$ the FP $L(\mathbf{q},\omega)$ can be arbitrary large in the $t \to 0$ limit. Within the Kubo formalism $\zeta$ and  $\alpha_{\lambda}$ are expressed in terms of the pressure-pressure correlation function. In the diagrammatic expansion of the correlation function close to $T_c$ diagrams containing $L(\mathbf{q},\omega)$ bring the dominant contribution. Based on the experience gained in condensed matter physics \cite{4} we assume that the Aslamazov-Larkin (AL) diagram shown in Fig.\ref{fig:FIG1} plays the leading role. In the short presentation we leave out the detailed evaluation of this diagram which may be found in \cite{16,17}.

The important point is that the imaginary part of the polarization operator corresponding to the AL diagram is proportional to $t^{-3/2}$. Then the Dyson equation yields 

\begin{equation}
\alpha_{\lambda} \simeq R\cdot\omega\cdot t^{-3/2}\cdot \ln^2\dfrac{\Lambda}{2\pi T_c},
\label{07}
\end{equation}

where $R$ has a dimension m$^{-1}$ and depends on the critical temperature, Fermi momentum and the quark mean free path \cite{16,17}, $\Lambda$ is the UV cut-off equal to the Debye frequency in BCS. The bulk viscosity is 

\begin{equation}
\zeta \simeq R\cdot\dfrac{\varepsilon c^2}{\pi}\cdot t^{-3/2}\cdot \ln^2\dfrac{\Lambda}{2\pi T_c}.
\label{08}
\end{equation}

The temperature dependence $t^{-3/2}$ is rather close to the scaling law (\ref{05}). Finally, we note that the electrical conductivity given by the AL diagram has the $t^{-1/2}$ dependence \cite{17}. We also note that a broad range of acoustic problems in QGP has been examined in \cite{18}.

The author was supported by a grant from the Russian Science Foundation project number 16-12-10414. The author expresses his gratitude to M.S. Lukashov who participated in the early stage of this work.   

\bibliographystyle{apsrev}

\end{document}